# Advantages and developments of Raman spectroscopy for electroceramics

Marco Deluca[1,4,*], Hailong Hu[2], Maxim N. Popov[1], Jürgen Spitaler[1], and Thomas Dieing[3]

[1]Materials Center Leoben Forschung GmbH, Roseggerstrasse 12, 8700 Leoben (Austria)

[2]WITec (Beijing) Scientific Technology Co., Ltd. Xiaoyun Road 36, 100027 Beijing (China)

[3]WITec Wissenschaftliche Instrumente und Technologie GmbH, Lise-Meitner-Str. 6, 89081 Ulm (Germany)

[4]Present address: Silicon Austria Labs GmbH, Sandgasse 34, 8010 Graz (Austria)

*Corresponding Author: marco.deluca@mcl.at

**Abstract**

Despite being applied with success in many fields of materials science, Raman spectroscopy is not yet determinant in the study of electroceramics. Recent experimental and theoretical developments, however, should increase the popularity of Raman spectroscopy in this class of materials. In this Review, we outline the fields of application of Raman spectroscopy and microscopy in various electroceramic systems, defining current key bottlenecks and explaining promising recent developments. We focus our attention to recent experimental developments, including coupling Raman spectroscopy with other methodologies, and modelling approaches involving both the model-based data interpretation and the ab initio calculation of realistic Raman spectra.

## 1. Introduction

Raman spectroscopy is a very popular non-destructive analytical method used in chemistry, biology, medicine, materials science and forensics. Its strength resides in the capability to achieve an accurate chemical and structural fingerprint of materials. Raman spectroscopy is based on the Raman effect, which is the inelastic scattering of incident visible monochromatic electromagnetic radiation by interaction with a crystal lattice or a molecule[1,2]. Although many excitation processes could lead to Raman scattering, in ionic/covalent solids scattering by optical phonons (i.e. quantized lattice vibrations) is the most relevant. In a Raman experiment, the incident radiation either gains or loses energy quanta by interacting with optical phonons. The resulting inelastically scattered radiation is constituted by so-called Anti-Stokes or Stokes lines (related to photon energy gain and loss, respectively). The process leading to Stokes lines originates from the ground state and is thus more probable at room temperature. Hence, Stokes lines have stronger intensity and are generally measured in Raman spectroscopy[3]. Owing to the physical mechanism at their basis, Raman spectra are sensitive to any change in the material that alters the energy of optical phonons, like the breaking or formation of chemical bonds, phase transitions, residual strain/stress, or the presence of lattice defects.

One of the advantages of Raman spectroscopy is the hyperspectral imaging capability[4], which can be achieved by coupling a Raman spectrometer with a microscope. The possibility to quickly obtain 2D spectral maps is a strong advantage of Raman spectroscopy over X-ray diffraction (XRD), where spatially resolved analyses can be carried out only by using expensive area detectors, and is generally restricted to synchrotron radiation facilities. Another aspect better highlights the complementarity of Raman spectroscopy with respect to XRD: The coherence length of Raman spectroscopy, defined by the product of the electron speed (~$10^6$ m s$^{-1}$) with the time the electron takes to experience a scattering event (~$10^{-15}$ s for Raman scattering), is in the order of nm (cf. Ref [5]), at least one order of magnitude shorter than XRD. This means that Raman spectroscopy is more sensitive to the local, short-range structure of materials.

Electroceramics are ionically or covalently bonded solids that demonstrate a specific functionality or property in response to an applied electric field, or produce an electrical response to another (e.g. mechanical, thermal) stimulus. Typical examples are dielectric (i.e. insulating), ferroelectric, piezoelectric, magnetoelectric/multiferroic, varistor, pyroelectric, thermistor and semiconducting ceramics. Such materials can be produced as bulk ceramics, single crystals, thick films (i.e. thickness above 1 µm) and thin films (thickness below 1 µm). Electroceramic thin films, moreover, can be produced by many different methods including physical/chemical vapour deposition, epitaxy, and chemical solution deposition, whereby the choice of deposition method impacts on the achievable microstructure. The structure-property relationships in electroceramics are often a multiscale problem, as the property originates from the local scale (e.g. cation displacements at unit cell scale in case of ferro/piezoelectrics) but the functionality is rooted on the mesoscale (e.g. the behaviour of ferroelectric domains) and the microscale (grain structure). The main issue for characterization is related to the local structure (< 10 unit cells), which may be different than the average one, and is often decisive for the properties of electroceramics. Raman spectroscopy may address successfully most of the challenges of electroceramics, including the effects of short-range order or disorder. In fact, it accesses short-range information and can perform spatially-resolved (hyperspectral) analyses, hence it can bridge length scales up to the microstructure level. Its short-range sensitivity can uncover structural information decisive to study structure-property relationships in electroceramics, and it does so in a table-top and easy-to-use setup, which is unthinkable for most laboratory techniques.

Although Raman spectroscopy has become very popular recently in materials science fields, its application to electroceramics appears to occur at a slower pace. From the over 45000 articles published from 2011 onwards on Raman spectroscopy in materials science, 50% dealt with carbon (nanotubes and graphene), 21% with semiconductor materials (including 2D materials but excluding graphene), 18% with polymers and nanocomposites, 9% on ceramics in general (excluding electroceramics) and only 2% with electroceramics (Source: Scopus). This number is somewhat surprising if we consider the recent popularity of electroceramic materials especially for piezoelectric and energy storage applications[6–11].

There might be, obviously, many reasons for this discrepancy in the usage of Raman spectroscopy in electroceramics compared to other materials science fields. One is certainly the fact that there are few research groups worldwide working on electroceramics, and in those groups the use of other characterization methods like XRD is preferred, because one can directly obtain structural parameters from them. In addition, Raman scattering is a weak phenomenon; hence, signal/noise ratio may be low in materials like electroceramics. This is true especially if the surface of the material is either highly refractive or absorbing, or rough, and in thin films, where the small interaction volume reduces the available signal[12]. In addition, the Raman signal is extremely material-dependent, and some electroceramics may not be very good scatterers.

But the most important reason for which many groups working with electroceramics do not consider Raman spectroscopy as a characterization method, is probably a perceived complexity in the interpretation of the Raman measurements, which involves:

- Uncertainties in the definition and operation of spectral fitting.
- Lack of knowledge of the Raman-active phonons in solid materials, especially electroceramics with complex chemical formula.
- The fact that, in materials with long-range electrostatic forces like electroceramics, identification of peak positions is hampered by the presence of oblique phonons (longitudinal optical/transverse optical, LO/TO, splitting)[13–15].
- The difficult interpretation of Raman spectra in defective or disordered crystalline materials, like many electroceramics are.
- The complex theory needed to study residual stresses in electroceramics.
- The perceived long acquisition time of hyperspectral Raman datasets.

Nowadays, however, new developments in the Raman equipment (confocal Raman microscopes, fast mapping stages) and – especially – new procedures and capabilities for the computation of Raman spectra both from first principles and on the mesoscale, can help to overcome these apparent limitations and lead to an increased use of Raman spectroscopy in the field of electroceramic materials in the near future.

In this article, we review numerous applications of Raman spectroscopy in various fields of electroceramics, at the same time presenting the latest experimental and theoretical advancements that can transform Raman spectroscopy into an easy-to-use and widespread characterization method for this class of materials. We start (Section 2) by describing Raman spectroscopy from the point of view of instrumentation, also outlining recent technological advances, and data analysis procedures. Then (Section 3) we outline possible applications of Raman spectroscopy to the study of electroceramics, including both state-of-the-art and cutting-edge studies. We continue in Section 4 by describing novel computational methods that can help the interpretation of Raman spectra in electroceramics, followed by a short discussion and summary. Although far from being exhaustive, this review aims to showcase a number of promising developments that will potentially increase Raman spectroscopy's applicability in this exciting materials field.

## 2. Raman in practice

*2.1 Raman instrumentation*

The equipment for Raman spectroscopy, in the much-used micro-backscattering configuration, employs an optical microscope to channel both the incident laser (used as a monochromatic excitation line) towards the sample and the Raman-scattered light towards the detector. In between, the core of the system is constituted by the spectrometer unit (also called spectrograph). It consists of a grating that disperses the scattered light onto the focal plane of a (multichannel) detector, for example a charge-coupled device (CCD). The light collected by the microscope, however, comprises also the elastically scattered Rayleigh spectral portion, which is much stronger (of a ~$10^7$ factor) than the inelastically scattered Raman signal, and thus must be filtered before reaching the detector[16]. There are many possibilities to filter out the Rayleigh line. Most spectrometers possess a single grating to disperse the scattered light, hence the Rayleigh scattering must be removed before reaching the spectrometer, which is done using a high-end edge or notch filter. An alternative is to use a triple monochromator, namely a spectrometer possessing three gratings, in which the first two work as a filter to remove the Rayleigh radiation. This system allows to cut the elastic scattering signal very precisely and thus to measure the spectrum almost down to the excitation wavelength, however the use of multiple gratings (and related optical equipment like mirrors) greatly suppresses the Raman signal, which is a drawback for weak scatterers like most electroceramics. Recently, new filter/grating systems were implemented on Raman spectrometers for Rayleigh light suppression. These systems are now state-of-the-art and allow a good suppression of the Rayleigh peak (typically > OD6, OD = optical density) while maintaining a high transmission of the Raman scattered light (typically > 90%)[3]. Before reaching the spectrometer, the collected light can be analysed in terms of its polarization (i.e. a polarization filter is placed on the scattered light path). Further, a confocal pinhole may be placed before or after the polarisation analyser to select scattered light only coming from the vicinity of the focal point on the sample (which improves the lateral and depth resolution). Novel modular systems involve connecting microscope and spectrometer with optical fibres. This way, by adjusting the fibre collimation it is possible to work always at the highest confocality, without the need for an additional confocal pinhole, since the core of the fibre acts as the pinhole itself.

The use of a microscope to focus and collect incident/scattered radiation brings about several advantages for in-situ measurements. Samples can be placed in cryostats or on heating stages (with variable temperature ranges, but in general measurements are possible from 4 K up to ~1800 K), with the possibility of concurrently applying electric or magnetic fields. Concerning the former, generally values up to 500 V can be reached, otherwise the measurement must be done in insulating oil to avoid arcing, which might affect the Raman signal. Also stress can be applied in-situ in tensile, compressive or bending configurations, which is necessary to determine the shift-stress correlation to enable residual stress analyses. High pressure studies are also popular (especially in the field of mineralogy), and they require a diamond anvil cell (DAC) to apply hydrostatic pressure in excess of several GPa. In such equipment, a ruby crystal is generally inserted in the DAC alongside the sample, and used as a stress sensor (due to ruby's well-known shift-stress correlation).

## 2.2 Recent advancements in Raman instrumentation

Before the turn of the century, Raman spectroscopy and microscopy were mostly accessible only to researchers being capable of building large parts of the setup by themselves on an optical table. Since then, the reliability, user friendliness and thus accessibility of Raman microscopes has tremendously improved and high-end confocal Raman microscopes with intuitive, user friendly and automated setups are available from various vendors. Such setups mostly allow well-controlled beam delivery in terms of power adjustment and reproducibility as well as polarization state of the incoming laser. In addition, control of the incoming laser polarization can be motorized for some of the setups, but care has to be taken to avoid negative effects on polarization maintenance if dielectric or dichroic mirrors are used within the setups (cf. Section 3.3). Video guided sample positioning as well as automated Raman imaging is generally available in such setups and automated polarizer/analyser series, where a set of spectra for different polarization and/or analyser states is automatically recorded, can also be available.

The excellent in-situ capabilities of Raman spectroscopy were recently enhanced by realizing the connection between Raman spectrometers and other advanced scientific equipment, such as atomic force microscopes (AFM) and scanning electron microscopes (SEM):

- For AFM, the solution is mostly the integration in a single microscope platform, where the change from one to the other technique can be as simple as the turn of the microscope turret (used for example in Ref. [17,18]). Solutions where previously registered sample holders are used and samples are transferred between different instruments are also available. The combination of AFM and confocal Raman imaging also shows significant potential for the enhanced understanding of electroceramic samples[19,20]. Using AFM, the fine structure of the sample surface can be measured and by electrostatic force microscopy (EFM), scanning capacitance microscopy (SCM) or Kelvin Probe Force Microscopy (KPFM) also electrical properties of samples can be probed. Correlating those results with the molecular, orientation and strain information obtainable by confocal Raman microscopy allows further insights into the nature of the samples.

- For Raman-SEM integrations two general setups are realized. One uses a parabolic mirror, which is positioned under the SEM column in much the same way as for the case of cathodoluminescence (CL) measurements. The challenge herewith is that in contrast to CL the resolution is defined by the laser/Raman optics and wavelength, whereas in CL it is determined by the electron beam. Thus, the resolution and throughput of such systems, which is determined in part by the achievable numerical aperture of the focussing optics, often suffer. An alternative solution uses a vacuum capable objective inside the SEM chamber but at an offset relative to the SEM column. This requires the sample to be moved by a sample positioner inside the chamber but allows for diffraction-limited resolution and high throughput comparable to standalone systems. Since the space around an SEM is always scarce, Raman systems with a small footprint and/or fibre coupled lasers and spectrometers are advantageous in such scenarios[21,22]. While it is established to use SEMs in combination with chemical analyses for characterization (see for example Ref. [23]), the combination with high resolution confocal Raman imaging inside the SEM chamber[24] is currently less used. It however shows a significant potential for the correlation of the fine structure as can be evaluated by SEM with the molecular information and strain state as revealed by confocal Raman imaging.

Further, the use of second-harmonic generation (SHG) imaging in combination with confocal Raman imaging is an upcoming field of research. V. Gopalan and his co-workers have applied polar SHG imaging to identify tetragonal $BaTiO_3$ films and their domains (90° or 180° domains). They determined also the existence of a monoclinic phase, which was further confirmed by polarised Raman Spectroscopy[25]. Similar analyses have been conducted on $KNbO_3$[26] and $CaTiO_3$[27] films. Research in other fields of material science has already made use of available systems combining SHG and Raman in one instrument[28,29]. However, the use of such combined systems in the field of electroceramics still seems to be limited.

### 2.3 Curve fitting

Curve fitting is probably the single most important procedure for data treatment in Raman spectroscopy. It involves using a mathematical expression to model the phonon modes in the spectrum and the use of a minimization routine to achieve the best-possible fit to the measured spectral signature. A typical workflow for spectral fitting is:

- Extraction of the portion of Raman spectrum to be fitted.
- Baseline correction, i.e. subtraction of a suitable function (linear, polynomial or spline/adaptive) to remove the background.
- Placing the first approximation of the phonon modes.
- Automatic iteration until the best-fit is obtained.

The most crucial aspects in this procedure are the baseline subtraction and the first approximation of phonon modes. Subtracting the wrong baseline might result in removing not only the background, but also portions of useful spectral information. The first approximation, on the other hand, must be based on physically sound grounds, namely by adding only the number of phonon modes that is permitted by theory. This is especially important in the treatment of Raman images, where the first approximation is applied generally to the whole dataset and not to single spectra. The user should prioritise physical soundness over the goodness of fit here.

The fitting parameters may vary depending on the peak function used, but in general there are three main parameters that can be extracted from fitting a phonon mode:

- **Peak position:** Is the "centre of mass" of the peak, and stands for the energy of the vibrational mode in question. As such, it depends on the strength of the atomic bonds related to that specific vibrational mode. It is influenced by any changes in the vibrational energy of the mode, caused for instance by changes in atomic mass, bond character (hybridization), electrostatic forces, spin forces, and strain.

- **Peak intensity:** Generally taken as the integrated intensity, it depends on the polarizability of the vibrational mode in question. It is influenced by the scattering geometry (hence by the crystal orientation), the presence of polar phases, and changes in phase amount.
- **Peak width:** Defined as the full width at half maximum (FWHM), it depends on the coherency of vibration, and is influenced by temperature and disorder. Very similarly to XRD diffraction peaks, a Raman peak will be broader at higher temperature and in presence of disorder. Crystalline samples have sharp Raman peaks. Amorphous or defective samples broad ones.

Once extracted, these parameters can be plotted against any control variable like composition, temperature, pressure, electric field etc., in order to visualize changes in material behaviour like phase transitions or any other structural modulation.

There are various choices concerning the peak functions that can be used for the fitting. Molecular vibrations are better reproduced by Lorentzian shapes, whereas phonons in solids (especially in electroceramics) should be modelled with a sum of damped harmonic oscillators (DHO), in analogy with the Drude-Lorenz model used in reflectivity spectra to fit phonon resonance peaks[30]. Ideally, a fitting program should allow not only multiple peak fitting (i.e. to model the spectrum as a whole), but also multiple choices of functions to be used in the same fit. Mixing DHO components with Debye relaxation functions, for example, allows to co-fit the central mode often appearing in ferroelectric or relaxor compounds[31]. Regrettably, there is still no ad-hoc software for Raman spectroscopy that has these characteristics. Built-in analytical software offered by Raman equipment producers may provide multiple peak fitting, sometimes with the availability of user-defined functions. Many Raman users, however, prefer self-programmed solutions in other mathematical software environments, including open source ones (python). Although this way they obtain a fitting program tailored to their needs, the time and effort needed to reach a reasonable working version of such a program must not be underestimated.

### 3. Application to electroceramics

*3.1 Phase transition studies*

In ferroelectric materials, phase transitions are often *displacive*, meaning that they involve the displacement of atoms or ions. Such transitions are underlined by the large frequency variation of an optical phonon, named "soft mode", which tends to zero (i.e. "softens") by approaching the phase transition[32]. Raman spectroscopy is thus very useful to determine the transition from paraelectric to ferroelectric phases by observing the shift of the soft mode at the Curie point[31,33,34]. This so-called "soft mode spectroscopy" requires measurements very close to the Rayleigh line, as these modes belong to low-wavenumber lattice vibrations (cf. Figure 1a-c). In contrast, "hard mode spectroscopy" is intended as the study of subtle variations of peak position, intensity and width for peaks not directly related to structural phase transitions, but that may be influenced by those[32]. Hard mode spectroscopy thus involves the observation of higher-wavenumber modes as a function of temperature or pressure[32,35–37] (cf. Figure 1d). Anomalies in the temperature evolution of hard modes can in principle be related also to second-order or order-disorder phase transitions, also very common in electroceramics. Notable examples of both soft- and hard mode spectroscopy are the study of the phase diagrams of $PbZrTiO_3$ [38–43], $PbMg_{2/3}Nb_{1/3}O_3$ (PMN) [44] (cf. Figure 1b-c), $PbMg_{2/3}Nb_{1/3}O_3$-$PbTiO_3$ (PMN-PT) [45,46], $Bi_{0.5}Na_{0.5}TiO_3$ [47–49] (cf. Figure 1d), $K_{0.5}Na_{0.5}NbO_3$ [50,51] (KNN), $BaTiO_3$ [52,53] (cf. Figure 1a) and solid solutions [54–66]. Those studies required high-end spectrometers, with triple monochromators to enable spatially filtering the Rayleigh signal, thereby approaching down to ~10 $cm^{-1}$ and below, where soft mode analysis can be carried out. Triple spectrometers, however, are very inefficient in terms of throughput, since they use a high number of gratings and mirrors, each one impacting negatively on the final obtained intensity (at the detector). Modern Raman spectrometers, on the other hand, are built with high-end elastic scattering filter systems allowing the detection of low wavenumber Raman bands down to ~10 $cm^{-1}$ even without using triple monochromators, while still effectively suppressing the Rayleigh peak. This experimental development makes soft mode and hard mode spectroscopy analyses easier and faster, with the prerequisite that cryostats and heating stages to precisely determine transition points are available. In the case of analyses under pressure, solid-state high-pressure equipment such as a Diamond Anvil Cell (DAC) is needed, where the sample and a pressure sensor (e.g. a ruby crystal) are inserted together in a cell filled with incompressible fluid and set under pressure. Typical examples are in-situ pressure-dependent studies carried out in relaxor perovskites [47,67,68], where the study of high-pressure phases helped understanding the role of chemically different nanoregions on the macroscopic dielectric behaviour. In another study, the substitution of Ba for Pb in perovskite-type $Pb_{0.78}Ba_{0.22}Sc_{0.5}Ta_{0.5}O$ relaxor ferroelectric over a broad pressure range has been observed [69].

In modern spectrometers, structural Raman characterization under different temperatures [70,71], pressures [69,70] or electric/magnetic fields [72,73] can be combined with confocal microscopes to attain in-situ imaging with resolution close to the diffraction limit (~300 nm lateral and ~1 μm in-depth, depending on laser wavelength and objective choice). Furthermore, confocal Raman imaging has also been demonstrated at cryogenic temperatures down to 2 K and with simultaneous applied magnetic fields up to 9 T [74,75]. This is reflected by recent work on the analysis of soft modes in rare-earth perovskite orthoferrites [76], Ba-based double perovskites [77], $LaAlO_3$ [78], $BiFeO_3$ (BFO) [79] and $AgNbO_3$-$LiTaO_3$ [80]. Other studies recently combined hard-mode Raman spectroscopy with IR studies to elucidate the phonon evolution and the phase transitions in the $K_{0.5}Na_{0.5}NbO_3$-$LiBiO_3$ solid solution [81]. The application of electric fields to multiferroic ceramics, while monitoring the changes in the Raman spectra, also proved to be of significant interest. Especially when not only observing single spectra, but rather confocal Raman images, changes in individual domains can be observed [82]. Furthermore, the combination of environmental chambers and/or electric field application with surface tracking techniques, where the sample is constantly kept in focus irrespective of its expansion or contraction, proves especially beneficial for these in-situ observations.

Ferroelectric and related materials have a broad spectral signature due to intrinsic translational disorder and oblique phonon behaviour [53]. This poses significant challenges for the interpretation of Raman-active modes. In polycrystalline ferroelectrics, the measured spectrum is the result of a volumetric averaging over an ensemble of crystallites with random orientations, hence mixed modes (named "quasimodes") - rather than well-localized peaks related to polar phonon modes - are observed. Consequently, the spectral signature is composed of broad convolutions of polar phonon modes, in which the contribution of each polar mode is impossible to resolve without information on the crystalline texture, i.e., the crystal orientation distribution, at the measured location[64]. Although texture analyses are possible with the aid of polarization filters and suitable computational routines (see Section 3.3), mode convolution in polycrystalline ferroelectrics affects the interpretation of phase transitions, as seen in the determination of the monoclinic phase of $PbZrTiO_3$ by Raman spectroscopy[41,43]. A possible workaround involves the study of the temperature evolution of the peak position modelled with the anharmonic softening behaviour from Balkanski, Wallis and Haro[83]. Any anomaly in the position vs. T diagram is visible as a deviation from the anharmonic softening behaviour, and can be ascribed to a phase transition. This method has been applied successfully in materials with broad and nearly featureless spectral convolutions, such as multiferroic ceramics[84–86], ferrites[87], and $Bi_{0.5}Na_{0.5}TiO_3$ ceramics[88,89].

*3.2 Disorder and defects*

Despite the difficulty in interpreting the broad Raman spectra of ferroelectrics and related materials, Raman spectroscopy has proven to be a very powerful method for studying defects and disorder effects in such materials, including transition to relaxor behaviour (cf. Figure 2a). Due to the aforementioned very short coherence length, Raman spectra can detect the presence of defects or structural variations on the nanometre range. The introduction of lattice defects (for example, by increased substituent contents in solid solutions) produces different vibration modes compared to the host lattice. This can be visualised by the appearance of extra Raman modes, as testified for example by B-site cation[90] and oxygen vacancy modes[91] in perovskites. In $BaZr_xTi_{1-x}O_3$, the relaxation of the crystal structure due to the introduction of the larger Zr cation can be seen in a progressively increased broadening of the Raman spectral signature, even at liquid $N_2$ temperatures[55]. Transition towards a full relaxor state is determined if a composition displays the same broad spectral signature over the whole investigated temperature range (from cryogenic temperatures up to above the Curie point of the parent ferroelectric material)[64] (cf. Figure 2a). The occurrence of clustering upon substitution can be evidenced from the splitting or shifting of Raman modes, the so-called two-mode behaviour[37], since clusters with different chemical composition will tend to vibrate differently, with a distinctive spectral signature, because of differences in the atomic mass.

The broadening of the phonon modes, the appearance of new modes or of mode splitting, is thus correlated to the onset of short-range structural variations. This is often related to the zone-folding effect, namely the disorder-induced activation, at the Brillouin zone centre, of originally non-zone-centre modes[92]. Indeed, the appearance of first-order Raman modes in nominal cubic or pseudocubic centrosymmetric structures confirms the presence of low-symmetry distortions (such as phase coexistence, octahedral tilting, etc.) at unit cell scale. This is evident both in relaxor perovskites and in pure $BaTiO_3$ above the Curie point[53], the latter being caused also by intrinsic lattice defects[93] (cf. Figure 2b). Even ferroelectric distortions above the Curie point were determined recently by Raman spectroscopy both in pure BaTiO$_3$ [94,95] and in Sr-modified BaTiO$_3$ [95]. These were ascribed to residual stress caused via either grain size effects or (intrinsic and extrinsic) lattice defects. Although the interpretation of defects and disorder effects in electroceramics is powerful already on a qualitative basis, it would benefit from insights from ab initio calculations (cf. Chapter 4) to avoid potential misinterpretations, particularly in defective and disordered materials. One notable case is Nb-substituted BaTiO$_3$, where an extra mode at ~830 cm$^{-1}$ was initially ascribed to vibrations involving Nb cations [63]. Recent investigations and theoretical calculations have confirmed, instead, that this mode is related to B-site cation vacancies [90,96].

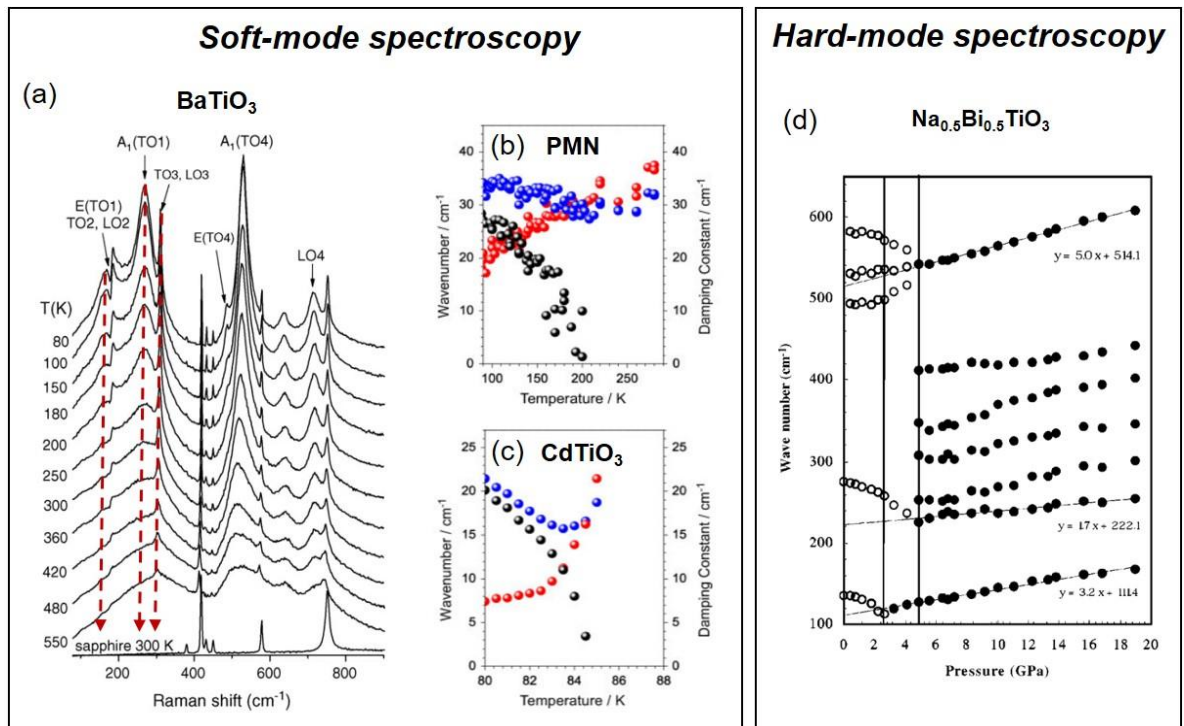

**Figure 1 - Examples of soft-mode and hard-mode spectroscopy, in electroceramics:** (a) Softening of phonons towards the Curie temperature in BaTiO$_3$ thin films [34]. © Elsevier (2005) – Reproduced with permission. (b), (c) Soft mode dynamics in PbMg$_{2/3}$Nb$_{1/3}$O$_3$ (PMN) relaxor ferroelectric and CdTiO$_3$ displacive ferroelectric, respectively. Blue, red and black circles denote the harmonic wavenumber, the damping constant, and the soft mode wavenumber. The results confirm that the phase transition of PMN retains displacive character while being under the influence of inhomogeneity (polar clusters). [31] ©John Wiley and Sons (2011) – Reproduced with permission. (d) Band position change as a function of pressure in the Raman spectra of Na$_{0.5}$Bi$_{0.5}$TiO$_3$. The lines are guides for the eye to emphasize spectral changes and equations are relative to a linear fit in the 5-19 GPa range. [48] ©American Physical Society (2001) – Reproduced with permission.

## *Disorder effects*

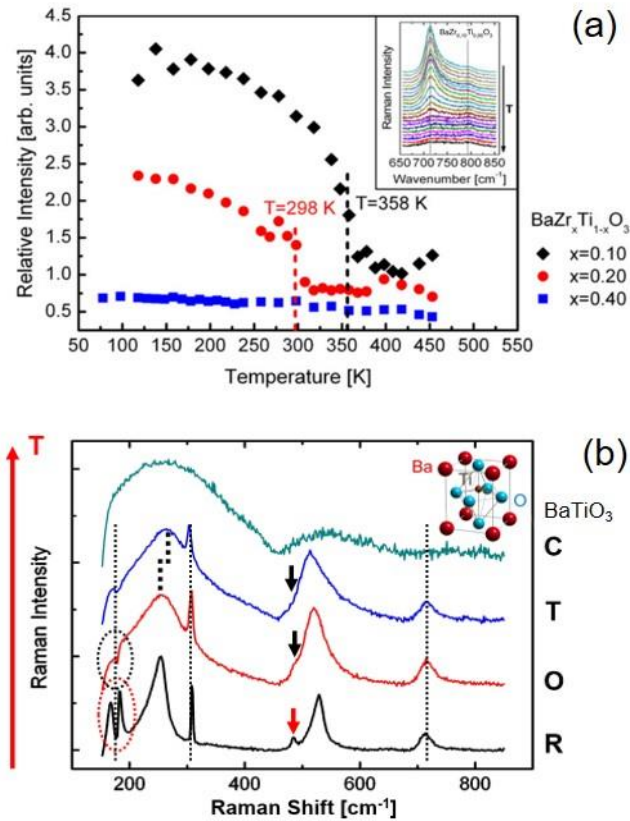

**Figure 2 - Disorder effects in electroceramics by Raman spectroscopy**: (a) Ferroelectric-to-relaxor crossover in $BaZr_xTi_{1-x}O_3$ ceramics. The Curie point and relaxor crossover is indicated by the intensity ratio of the peaks at 715 $cm^{-1}$ and 780 $cm^{-1}$. Samples with x = 0.10 and 0.20 show an abrupt relative intensity drop indicative of a ferroelectric-to-paraelectric transition. For x = 0.40, no transition is detected, which is typical of systems with relaxor character.[64] ©IOP (2014) – Reproduced with permission. (b) Raman spectra of pure, polycrystalline $BaTiO_3$ ceramics in the rhombohedral (R), orthorhombic (O), tetragonal (T) and cubic (C) phases. Circles indicate low-frequency modes that become damped in the O phase; arrows indicate the mode at 490 $cm^{-1}$ that disappears in the tetragonal phase. The appearance of a Raman spectrum in the C phase is a result of local spatial disorder.[53] ©John Wiley and Sons (2015) – Reproduced with permission.

Raman spectroscopy has been successfully used in electroceramics also for spatially resolved phase analyses, such as polytype detection in SiC[97,98], solid solution studies in AlN-SiC composites[99], and the determination of heterophase mixtures in the relaxor ferroelectric $Pb(In_{1/2}Nb_{1/2})O_3$-$Pb(Mg_{1/3}Nb_{2/3})O_3$-$PbTiO_3$[100]. Quantitative analyses of phase segregation, similar to the determination of monoclinic fraction in zirconia ceramics[101,102], however, require the presence of suitable standards with a known phase fraction[103]. Such standards are generally hard to establish in electroceramics.

### 3.3 Texture analyses
Due to the crystalline nature of many electroceramic materials, polarisation dependent Raman studies can in addition shed light on the strain[104] and orientation[19] across whole samples or even individual grains[105], and also determine the symmetry of selected Raman modes[106]. Polarised Raman spectroscopy has been often used to determine ferroelectric domain orientation. The cross section of Raman scattering depends on the Raman tensor, whose shape is related to the underlying crystal structure[107]. Therefore, when a crystal is rotated under polarized light, the Raman scattering intensity (depending on the value of the Raman tensor components) can assume a sinusoidal shape as a function of the rotation angle. The shape of the intensity sinusoid, in this case, depends on the symmetry of the crystal and the symmetry of the considered Raman mode[107]. Consequently, in uniaxial crystals like ferroelectrics the angle between the crystalline polar axis and the incident light polarisation can be determined by analysing this sinusoidal dependence. This is possible also in polycrystalline materials, provided that a preferential (non-uniform) texture (like in a poled ferroelectric) exists[108,109] (cf. Figure 3). In this case, in the measured location, the amplitude of the sinusoid bears information on the texture degree, which can be expressed in terms of an orientation distribution function [110]. Determinations of the texture degree from the measured sinusoidal intensity profile are generally carried out pointwise assuming a suitable domain distribution function[111]. Röhrig et al.[112] showed that common mathematical formulations of distribution functions are unable to effectively reproduce Raman data collected on ferroelectric polycrystals. On the other hand, they showed that using a Reverse Monte Carlo (RMC) model to calculate the sinusoidal intensity profile from a discrete texture representation, the closest match to the underlying ferroelectric domain texture can be attained[112]. This model was used to analyse ferroelectric domains in unpoled, poled and mechanically compressed $PbZrTiO_3$ polycrystals, and also in commercial piezoceramic actuators subjected to electric field and compressive stress[113] (cf. Figure 3). Previous ferroelectric domain mapping always attempted measuring the preferential domain direction upon the assumption of a constant domain texture over the whole map[114]. To determine the texture in each point of the map, in fact, the sample would have to be rotated exactly around the optical axis of the microscope in each point. This type of measurement is generally not possible in common Raman setups (unless the instrument allows the free rotation of the incoming polarization vector along with the analyser). These analyses, thus, have to be considered as a qualitative mapping of relative ferroelectric domain orientation. Both 2D and 3D domain orientation mapping were carried out this way, whereby 3D analyses were restricted to single crystals to take advantage of the material's transparency[115]. These studies involved using a lateral and in-depth probe response function to avoid effects related to probe convolution, however they failed to properly include the effect of the refractive index, which especially in semi-transparent materials can produce a strong compression of the depth of focus due to spherical aberration[4]. The only way to minimize this issue is the use of an oil immersion objective to reduce the refractive index difference between outside and inside the measured medium[4]. This strategy, however, cannot be applied to electroceramics, since no immersion oil exists with a refractive index as high as those of electroceramic materials. The best approach for 3D domain mapping is thus to use a confocal Raman microscopy system, that can provide highest confocality (i.e. < 1 µm) while maintaining a high throughput to allow for sensible measurement times. For example, F. Rubio-Marcos et al. have clearly distinguished a-c-domain walls and their b-domain boundary in the depth Raman imaging of a $BaTiO_3$ film (cf. Figure 4)[20], even succeeding to influence domain wall motion via changing the laser polarisation[82]. It should be noted that the focus distortions due to the large refractive index difference between the sample and the air had to be taken into consideration also in this case. Confocal Raman imaging has also advantages for the other aforementioned analyses. For instance, owing to the high spatial resolution, researchers have successfully resolved the polymorphic phase boundaries[116], their size, and the location of submicron secondary phases[27] in KNN-based piezoceramics.

Modern confocal Raman systems also allow the free rotation of the incoming polarisation as well as the free selection of the direction of the polarised light to be analysed. Some systems even allow for the motorization of these elements, which further facilitates the collection of a full polarisation dependent data set[27]. This type of setup, combined with a high confocality, has the potential of measuring submicron 3D ferroelectric domain orientation and texture maps. It should be noted, however, that in this case a calibration to avoid aberrations related to the polarization dependence of mirrors (dielectric mirrors in particular) and gratings is necessary. Optical fibre-coupled instruments bear a great advantage in this respect, as they minimise the use of mirrors in the optical path.

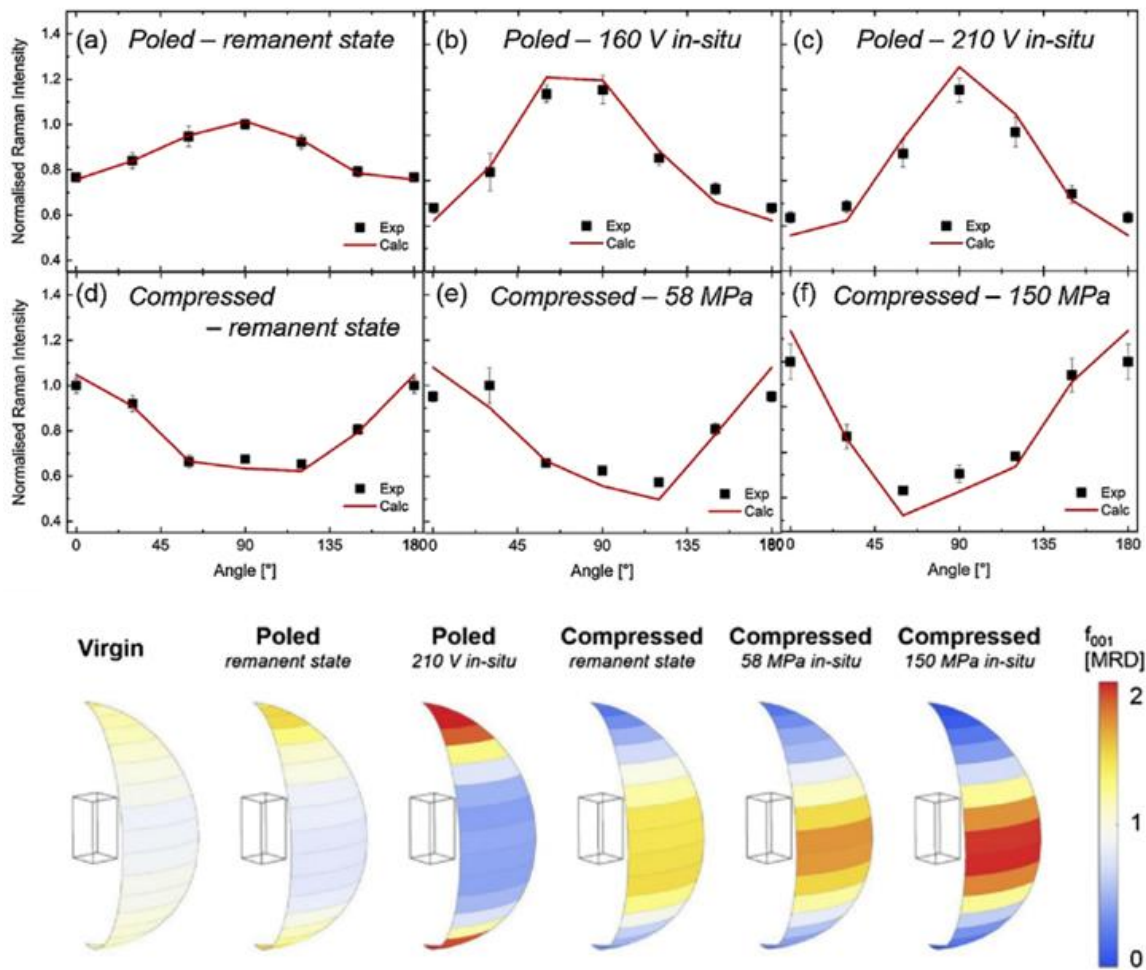

**Figure 3 – Ferroelectric domain texture analysis by Raman spectroscopy:** (a-f) Angular dependence of the polarised Raman intensity of the 220 cm$^{-1}$ phonon mode of PZT for samples measured in different states of electrical poling or mechanical compression along the axis parallel to incident laser polarisation. The solid line displays the best-fit obtained with the Reverse Monte Carlo procedure for domain texture analysis. Ferroelectric domain pole figures on the bottom displaying the ferroelectric domain texture distribution associated with the above angular dependence curves.[113]

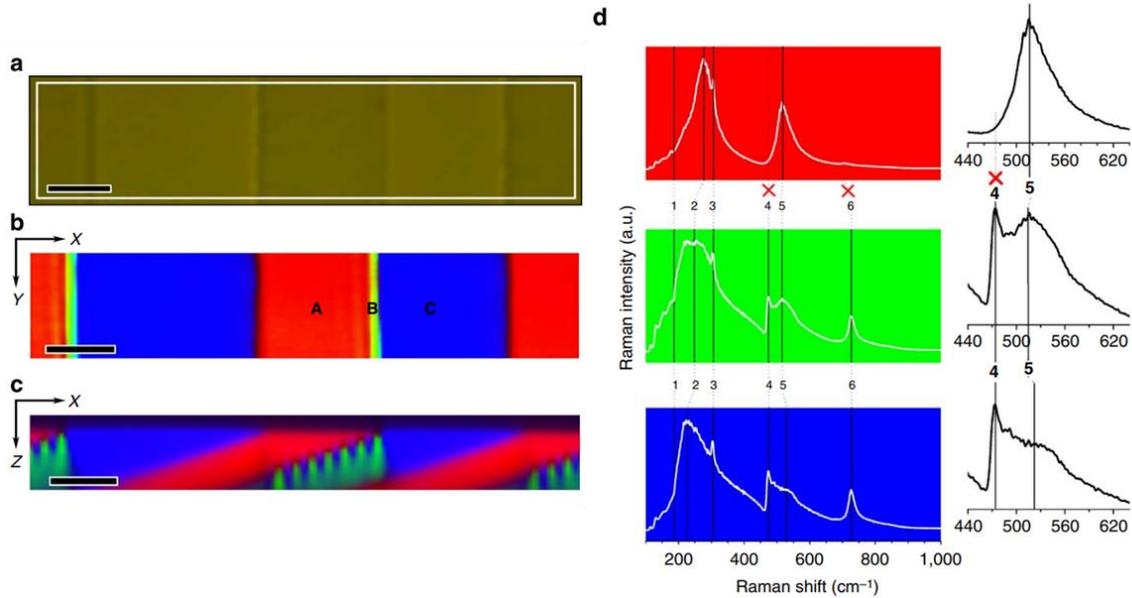

**Figure 4 - Mapping of the domain structure in BaTiO₃ (BTO) single crystal through confocal Raman microscopy:** (a) Optical micrograph of the area of the BTO single crystal subjected to XY and XZ Raman scans (white rectangle). (b) XY Raman scan showing the in-plane domain distribution. (c) XZ Raman scan showing the in-depth domain distribution. In (b) and (c), differently oriented domains are shown with different colours. The intensity of the colour is correlated with the Raman intensity. The dimension of the scale bar in (a), (b) and (c) is 20 μm. (d) Shows Raman spectra of BTO associated with the different colours in (b) and (c). The domains are distinguished by changes in Raman spectral shift and the relative intensity of Raman modes (marked by numbers in (d)): red = *a*-domain, blue = *c*-domain, green = *b*-domain (cf. points A, C, and B in (b)). The insets show magnified Raman spectra in the vicinity of the 4 and 5 Raman modes.[20]

### 3.4 Residual stress analyses

Raman spectroscopy is popular also for residual stress analyses in semiconductor and ceramic materials. The physical mechanism at the basis of this measurement is the change of the vibrational energy of the lattice (or of the spring constant of atomic bonds, in classical terms) in presence of strain. The theoretical framework is based on a semiclassical approach involving the so-called phonon deformation potentials[117,118], which are the first derivatives of the bond force constant tensor with respect to the applied strain. This formalism requires knowledge of phonon degeneracy (i.e. splitting of phonon modes in presence of strain) and wavevector (i.e. the direction of the phonon with respect to the crystal's principal axes). The latter aspects require carrying out polarised Raman analyses. The measurement of the full residual stress tensor is possible only in limited cases, such as single-crystalline semiconductors (Si[117], Ge[119], GaN[120], diamond[121], among others), for which also sets of phonon deformation potentials are known. Even in those materials, nonzero components of the stress tensor have to be assumed *a priori* to enable the analysis. Recently, an alternative, inverse approach involving the calculation of Raman shifts via strains simulated by the Finite Elements Method (FEM) proved to be successful – especially for complex structures[122–124]. In essence, this method compares experimentally measured Raman shifts with the ones calculated from FEM-simulated strain values, hence it serves as a validation of the FEM's method assumptions. This approach is more reliable than comparing measured and calculated stresses, for which also assumptions in converting Raman shifts into stresses are necessary. Another approach is piezospectroscopy, where the Raman shift relative to a zero-stress condition is directly related to the residual stress tensor[125]. The proportionality constants (called piezospectroscopic coefficients) have to be measured along the various crystal axes by applying a controlled stress field to the material. This approach also involves assumptions on stress tensor components and polarised Raman measurements, but proved to be effective if applied to either luminescence peaks[126–128] or in polycrystalline ceramics with known texture[129]. Notable – in the field of electroceramics – is the analysis of residual stresses in $Al_2O_3$-based Low Temperature Cofired Ceramics (LTCC)[130] and polycrystalline AlN thin films[129]. In the case of ferroelectrics, on the other hand, precise measurements of stress-sensitive phonon modes are complicated by the overall broad spectral signature caused by the LO/TO splitting and oblique phonon behaviour, which poses significant challenges for quantitative residual stress analysis. In essence, in ferroelectrics the peak shift is related not only to residual stress, but also to the angles between crystal polar axis and incoming/scattered light at the measured point. The peak shift relative to residual stress can thus be obtained after a correction procedure that takes into account the local crystalline texture, by identifying the crystal main axis in each investigated point, and assigning the corresponding piezospectroscopic coefficient[131]. This procedure, however, is cumbersome and can be carried out only in modern Raman spectrometers where the polarisation filters – rather than the sample stage – are rotated, especially if hyperspectral residual stress images are sought for. Peak shift maps in ferroelectrics, thus, should not be understood as residual stress maps, and have little sense if they are not accompanied by texture analyses on each measured point. It can be expected that these aspects will be tackled in the future using modern high-end Raman spectrometers, enabling concurrent texture analysis (i.e. with freely rotating polarisers and analysers) and automatic peak shift correction to attain residual stress values in real-time.

## 4. Computational procedures for Raman applied to electroceramics

### 4.1 Ab-initio calculation of Raman spectra in perfect and defective structures

Raman spectra can be calculated out of a model crystal structure using ab-initio methods. To this end, one has to compute the positions of the peaks, their width, and intensities. As it was already mentioned in the introduction, the peak positions are determined by the frequencies of the phonons involved in Raman scattering, with their wavevector close to the $\Gamma$-point. When the anharmonicity of phonons is considered, also the peak widths can be derived from the phonon lifetimes, though, this route is more computationally demanding and often the peak width is treated as a fitting parameter. Finally, the peak intensities are a function of the Raman susceptibility tensors (i.e. the Raman tensors), and the scattering geometry[132]. The simulated Raman spectrum of the considered single crystal and the scattering geometry is then obtained as a superposition of the peak shape-functions, e.g. Lorentzians, centred at the phonon frequencies. The required ingredients, i.e., phonons and Raman tensors, can be computed by most of the major DFT codes (e.g., VASP[134], Abinit[135], Quantum Espresso[136], CASTEP[137], CRYSTAL[138,139], etc.) and also by stand-alone packages (Phonopy[140], Phonon[141],

PHON[142], etc.). The ab initio calculations are often combined with the group-theoretical analysis, since the Raman tensors must obey the crystal symmetries. With the help of the Bilbao crystallographic server[133], one can define the number of Raman-active modes, their symmetry, and the scattering geometry enabling their appearance, all based solely on the crystal structure. However, purely symmetry-based analysis does not provide any information on the peak position and hence modelling is indispensable. The ab initio approach has been used, e.g., for predicting peak positions and Raman intensities of BeO, yielding a good agreement to the experimental data (see Figure 5a).

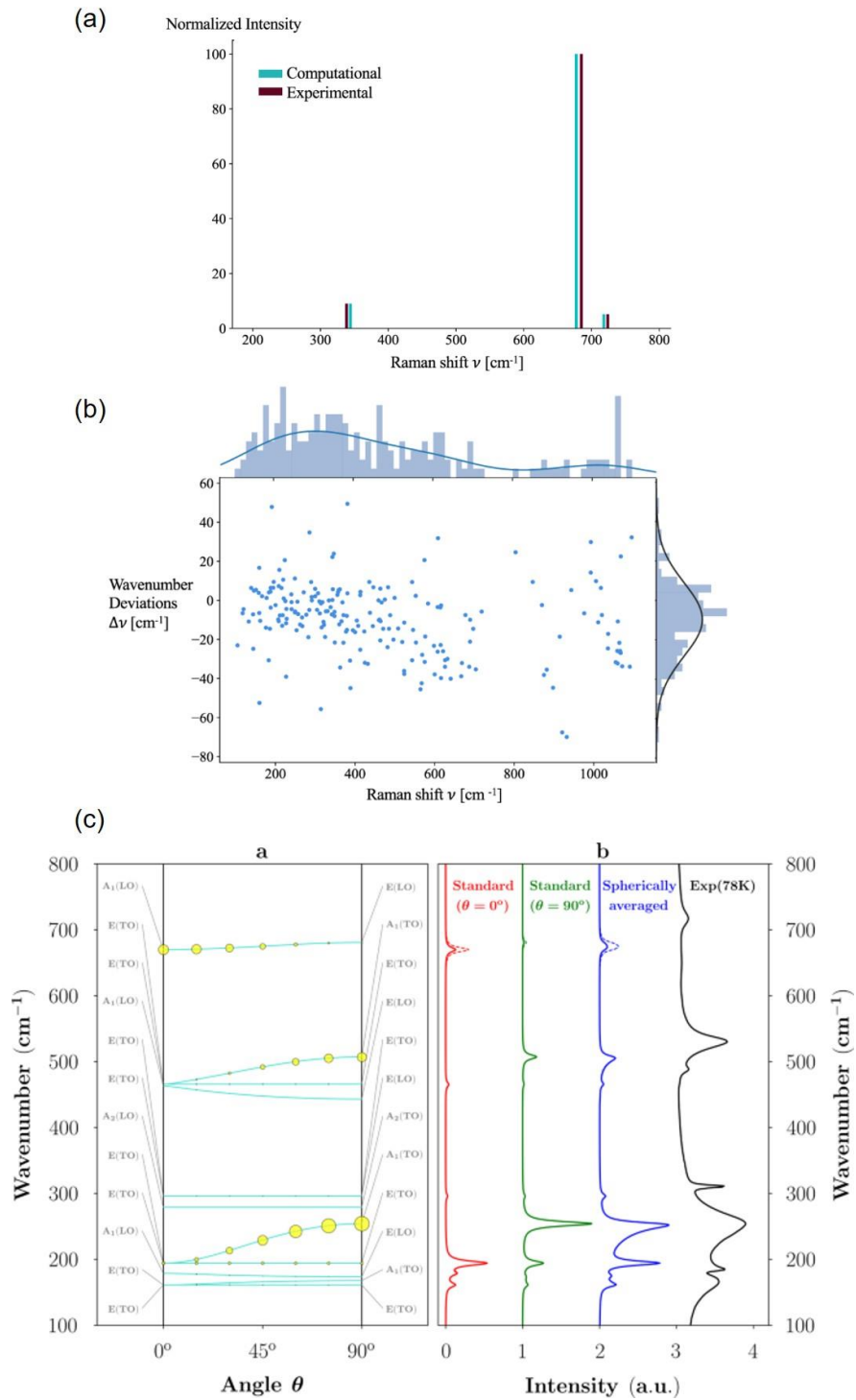

**Figure 5 – Raman spectra from first principles:** (a) Comparison between computational (ab-initio) and experimental Raman spectra of BeO material (Bromellite).[143] (b) Scatter plot consisting of wavenumber deviations $\Delta\upsilon$ of calculated modes vs. the position (in wavenumbers) of the corresponding (experimental) Raman modes. In other words, the y axis shows how much the wavenumbers ($\nu = \nu_{mode}$) of calculated phonon modes differ from those of their matched experimental peaks, namely $\Delta\nu = \nu_{mode} - \nu_{peak}$. The top x axis and the right-side y axis show the Gaussian kernel density estimates of $\nu$ and the normal distribution of $\Delta\nu$, respectively.[143] (c) a - Angular dispersion of the phonon modes as calculated from first principles ($\theta$ is the angle between the $C_3$ axis and the phonon q-vector); the dot size is proportional to the intensity of the Raman mode computed according to Eq. (7) in Popov et al.[144]. b – Comparison between calculated Raman spectra of polycrystalline r-BaTiO$_3$ (two different orientations – 0° and 90° - and spherically-averaged) and the experimental spectrum of BaTiO$_3$ measured at 78 K. Dashed lines indicate spectra calculated by omitting the electro-optic contribution.[144]

Theoretical Raman spectroscopy helps to assign experimentally observed modes and allows predicting Raman spectra that have not yet been measured. Results have been reported for simple substances - $\gamma$-Al$_2$O$_3$[145] and h-ZnO[146]; minerals - dolomitic lime[147], silicates[148]; transparent conductive oxides (TCO) - SrSnO$_3$[149], CaSnO$_3$[150]; non-linear optic (NLO) materials - LiTaO$_3$ and LiNbO$_3$[151], Bi$_{12}$SiO$_{20}$[152], ZnTiO$_3$[153], KSbGe$_3$O$_9$[154]; potassium antimonyl silicate (relaxor)[155]; multiferroics - BiFeO$_3$[156], RAlGeO$_5$ (R=Y, Sm-Lu)[157]; and photovoltaic double perovskite Cs$_2$AgSbCl$_6$[158]. Theory offers rich insights into the connection between the Raman signal and the stress state of the material. Here, a number of materials have been investigated theoretically: t-ZrO$_2$ under pressure[159], Aurivillius Bi$_2$WO$_6$[160], environmental barrier coatings for high-temperature applications (Yb$_2$O$_3$, Yb$_2$SiO$_5$, Yb$_2$Si$_2$O$_7$)[161]. Since Raman spectra are sensitive to the local ordering of atoms, Raman spectroscopy is widely used to reveal the ionic ordering/disorder in ionic conductors and substituted systems – Li$_{0.30}$La$_{0.57}$TiO$_3$[162], MgTa$_2$O$_5$[163], K$_{0.5}$Bi$_{0.5}$TiO$_3$[164], Na$_4$P$_2$S$_6$[165], vanadates MV$_2$O$_5$ (M=Na, Ca, Mg)[166–169] and metavanadates MV$_2$O$_6$ (M=Zn and Cu)[170], and Fe$_2$O$_3$[171] disordered by femtosecond laser irradiation. Other applications include clarifying the crystal growth mechanism of Li$_2$Mo$_3$O$_{10}$[172] and distinguishing polymorphs of HfO$_2$[173] and ZrO$_2$[174]. In these studies, the fidelity of simulated Raman spectra varies when compared to the experiment. Though theoretically predicted peak positions or even simulated Raman spectra provide a lot of insight, the nearly perfect match presented in Figure 5a is rather an exception.

The following discrepancies are typically observed:

a) Experimental spectra have more features (peaks and their parameters) as compared to the simulated ones;
b) theoretical peak positions often deviate from the experiment (see Fig. 5b);
c) the experimentally obtained peaks sometimes exhibit asymmetry, which cannot be described by a symmetric peak shape function like a Lorentzian.

The first problem stems from overly simplified structural models used in the simulation. For instance, the real material can have chemical, structural, and polar disorder. Unfortunately, it is not always possible to simulate such a disorder in atomistic simulations by increasing the complexity, and hence the size, of the employed structural model due to computational constraints. The second problem is likely due to the approximate exchange-correlation functionals used in the theoretical lattice dynamics. The third one can originate from Fano resonance[175–178], phonon confinement[179,180], oblique phonons[144], and the deviation of the material from a perfect single crystal. Indeed, any defect, introduced into a single crystal, breaks the translational symmetry, and can lead to the appearance of localized vibrations or activate non-Γ-point phonons due to Brillouin zone folding. Also, here, progress was recently made. Computed Raman spectra have been reported for a study of Ta-doping in h-ZnO[146], for point defects in SiC[181,182], for vacancies in $BaTiO_3$ and $SrTiO_3$[183], and defect clusters in Zr- and Nb-doped $BaTiO_3$[96]. It must be also considered that real materials are often polycrystalline, which poses additional complications that are addressed in the next section.

*4.2 Spherical Averaging for the calculation of Raman spectra in polycrystals.*

Polycrystals, powders, and ceramics, are essentially a collection of small crystallites or grains. Due to the finite size of the Raman experimental probe volume (typically in the range of 1-10 $\mu m^3$), the Raman spectra of numerous grains are collected simultaneously, while each grain of this ensemble can be oriented in an arbitrary way. Thus, simulating Raman spectra of fine-grained materials relies upon the methods developed for single crystals supplemented with averaging techniques. One must assume equal probability of every possible grain orientation, for instance using Placzek rotation invariants of the Raman tensors that are derived from the Raman tensors[184–186]. This approach is used, e.g., in the WURM project[148,187], which hosts a number of simulated Raman spectra, including of electroceramics. The same approach is used in a high-throughput study recently introduced by the Materials Project team[143], and yet another online database[188]. The limitations of this standard approach are relatively well understood:

a) it performs poorly in the case of polar materials (including electroceramics), which exhibit a strong directional dependence of the phonon frequency that affects the peak position. In addition, the Raman tensor itself is affected by the phonon propagation direction due to coupling with the electro-optic tensor, which renders the Placzek invariants incorrect;

b) Real polycrystals can have a texture.

The former limitation is addressed in the spherical averaging method proposed by Popov et al.[144], which introduces an averaging scheme that correctly takes into account the features of polar materials. This method has been shown to dramatically improve the simulated spectra of AlN, rhombohedral $BaTiO_3$, and $LiNbO_3$, resulting in an unprecedented match to the experiment. Figure 5c shows the computed Raman r-$BaTiO_3$ and also reveals the connection between the spectral features and the oblique phonons. This method was used successfully by Veerapandiyan et al. to elucidate the presence of cation vacancies in heterovalent-substituted $BaTiO_3$[96]. The issue of the texture still has to be addressed, although a combination between the spherical averaging method with existing texture models[112,113] might constitute a valid starting point.

*4.3 Raman hyperspectral imaging and related data treatment*

The combination of state-of-the-art Raman spectroscopy with confocal optical microscopy, enabled by the experimental advances described in Chapters 2-3, allows the collection of "hyperspectral" datasets, constituted for instance by 2D and 3D spatially resolved maps, in which each "pixel" corresponds to a Raman spectrum. Such hyperspectral Raman imaging requires also advanced signal processing to be able to extract the desired properties of a Raman signal in an automated way. The acquired data, in fact, need to be corrected with respect to undesired spectral features like wavelength-dependent instrumental sensitivity or background contributions (e.g. fluorescence of the sample or black body radiation at high temperatures)[189]. Then, the information of interest has to be extracted from the signal in an automated way. Consider for example phase quantification, a typical application that may involve hyperspectral imaging (i.e. the quantification of phases with different chemical composition[189] or crystal structure[190]). In this case, a straightforward way of analysing the Raman spectra is to assume that the spectrum of a polyphase material is a linear mixture of the spectra of the pure phases[189]. Then, a classical least-square (CLS) fitting procedure can be used to determine the phase proportions at each pixel. In recent years, however, signal processing based on machine-learning techniques has been used more and more for the efficient interpretation of Raman spectra in the generation of Raman hyperspectral images. One prominent example is the principal component analysis (PCA), where a spectrum is decomposed into an unknown number of orthogonal spectral sources (the principal components). Applying mathematical procedures, spectral-like information by order of decreasing importance is obtained, where only the first sets of data contain relevant information, while the following ones contain nothing but noise-like features. This allows rapid efficient extraction of information from the whole dataset (a spectrum at each (x,y) point of the map)[191]. PCA has also been used to recognize orthorhombic, tetragonal, and cubic phases as well as to construct the phase diagram in ferroelectric crystals[190]. The downside of PCA is that the principal components are by definition also allowed to have negative components and thus cannot be representative of real spectra. Methods such as the Non-Negative Matrix Factorization (NMF) can overcome these limitations to some degree[192]. Other flavours of machine learning approaches used for Raman hyperspectral imaging include e.g. Multivariate Curve Resolution Alternating Least Squares (MCR-ALS)[193].

While Raman hyperspectral imaging has been applied for more than fifteen years in areas like pharmacy and biology, this technique was applied to ceramics only rather recently, with a limited number of works published so far[194]. Maslova et al. used this technique to investigate the microstructure of a ceramic sample of uranium dioxide, including grain size, changes in local grain orientation, local strain and oxygen stoichiometry[191]. Hauke et al. applied Raman hyperspectral imaging to the in-situ investigation of densification and grain growth in silicate ceramics, including the appearance of metastable phases, by taking Raman hyperspectral images during heating and cooling[189]. In the thriving field of battery research, automatic Raman hyperspectral imaging has been used to analyse Lithium-ion batteries electrode materials[193]. In this work, the cathode material was investigated in terms of carbon and $LiMO_2$ (M = Ni, Mn, Co) content. Lambert et al. applied confocal Raman hyperspectral imaging to a ceramic-resin cement junction, where it was used to visualize the distributions of different compound concentrations[195]. Goj et al. determined the main crystalline phases in a glass-ceramic composite, including features which were not visible in XRD (like an $AlPO_4$ layer)[196]. Altogether, Raman hyperspectral imaging for ceramics and electroceramics is still a very young, strongly evolving field, and makes use of various approaches, so that it is too early to highlight best-practices. Nevertheless, given the quick development of the involved techniques it promises to rapidly gain importance and define well-established procedures within the next few years.

## 5. Outlook

Recent advancements in the Raman experimental capabilities, in the theoretical description of Raman spectra, and the increased availability of software and facilities for high-throughput calculations offer numerous possibilities for the Raman-based analysis of electroceramics. In particular:

- The use of confocal imaging equipment and motorized polarisation analysers allows 2D and 3D hyperspectral ferroelectric domain orientation and texture analyses with resolution close to the diffraction limit.
- The use of advanced elastic scattering filtering systems eliminates the need to use low-throughput triple monochromators to access ferroelectric soft modes.
- The combination of Raman equipment with stages allowing multiphysical characterizations, such as the concurrent application of temperature, electric and magnetic field, and pressure, open up interesting experimental possibilities for electroceramics.
- The development of novel correlative microscopy techniques (i.e. Raman + AFM, Raman + SEM, Raman + SHG) offer new opportunities for the study of electroceramics.
- Recently developed computational methodologies greatly aid the interpretation of Raman spectra in electroceramics, even contributing to a full automation of Raman analysis and data interpretation. These include the ab-initio calculation of Raman spectra in defective or disordered structures, the development of the Spherical Averaging method to calculate Raman spectra in *all* kinds of polycrystals (including polar ceramics), the use of Reverse Monte Carlo or Finite Element Methods to interpret Raman-based datasets for texture or residual stress analysis, and automated analytical routines for hyperspectral imaging (e.g. involving Machine Learning).
- Concerning the Raman data treatment and analysis, up-to-date Raman software packages allow user-friendly evaluation of full confocal Raman imaging data sets using single as well as multivariate analysis methods[197]. For the determination of exact peak positions to evaluate the strain state or of peak widths, which can be a measure for the homogeneity within the sampled volume, fitting algorithms allowing also user programmable peak functions have proven to largely facilitate the evaluation of Raman data sets. For the identification of the various phases, extended databases are available nowadays, as well as database management systems that can handle thousands of recorded spectra (i.e. full 2D confocal Raman imaging data sets) and compare them to databases or even determine the contributions of mixed phases.

The short-range order/disorder and multiscale structure-property relationships of electroceramics are accessible by Raman spectroscopy, but traditionally posed significant challenges for the interpretation of Raman measurements. Difficulties involving the presence of LO/TO splitting, the onset of extra-modes (and the related spectral broadening) in presence of defects and disorder, and in general the uncertainty in the assignment of number and symmetry of Raman-active modes, are tackled by the recent theoretical and computational developments of Raman spectroscopy. This, together with the availability of automated multiphysical experimental setups, will likely lead soon to a "coming of age" of Raman spectroscopy for electroceramics. In the near future, computationally-supported hyperspectral Raman datasets will enable 2D/3D quantitative defect, phase, texture and residual stress mapping. Further, correlative microscopy will enable using Raman as a chemical and stress sensor within complex structures prepared by Focused Ion Beam (FIB) in an SEM – for instance analysing degradation in layered electroceramic structures. The ultimate dream-scenario would be to combine Raman measurement software with databases of ab-initio calculated spectra, where the latter are obtained in real-time from atomic structures constructed by the user. This way, the effect of short-range order/disorder or defects would be made immediately visible and comparable with the experimentally measured spectra (for instance, at different conditions of electric field, temperature, pressure…), thus enabling obtaining quantitative information on the measured material. This approach would require enormous calculation power, thus direct connection to high performance computing (HPC) resources and access to ab-initio simulation software packages are needed. At the moment this may remain just an intriguing idea. It could however soon become reality, owing to the increase of HPC resources worldwide and the availability of novel non-Von Neumann computing approaches, which are likely to appear in the next decades.

## Acknowledgements

This project has received funding from the European Research Council (ERC) under the European Union's Horizon2020 research and innovation program (Grant Agreement No. 817190).

## Author Contribution

M. D. conceived the manuscript, prepared the initial draft, wrote the Introduction, largely contributed to sections 2, 3, and 5, and supervised the manuscript's preparation and revision; M. N. P. wrote sections 4.1 and 4.2; J. S. wrote section 4.3 and contributed to sections 4.4 and 3.2; T. D. and H. H. contributed to sections 2, 3, 4.3 and 5; all authors equally contributed to revising the manuscript.

## Competing interests

The authors declare no competing interests.

**Thorough introduction and review of piezospectroscopy.**

**throughput computation of Raman spectra.**